\documentclass[12pt,preprint]{emulateapj}

\usepackage{amsmath,natbib,graphicx}
\usepackage{epsf}
\usepackage{epsfig}
\DeclareGraphicsExtensions{.jpg,.pdf,.png,.eps,.ps}

\begin{document}

\title{The Statistical Significance of the ``Dark Flow''}

\author{
Ryan Keisler
}

\affil{Department of Physics,
University of Chicago, Chicago, IL 60637}
\affil{Kavli Institute for Cosmological Physics,
Chicago, IL 60637}

\email{rkeisler@uchicago.edu}

\begin{abstract}
We revisit recent claims of a significant detection of a bulk flow of distant galaxy clusters.  We do not find a statistically significant detection of a bulk flow.  Instead we find that CMB correlations between the 8 WMAP channels used in this analysis decrease the inferred significance of the detection to 0.7$\sigma$.
\end{abstract}


\bigskip\bigskip

\section{Introduction}
\label{sec:intro}

A recent set of papers \citep{kashlinsky08, kashlinsky09} claims to have detected the velocities of galaxy clusters with respect to the cosmic microwave background (CMB) frame by means of the kinetic Sunyaev-Zel'dovich (kSZ) effect \citep{sunyaev72}.  The papers suggest the existence of a ``dark flow'': a 700 km s$^{-1}$ bulk flow of all matter out to a redshift of at least $z\simeq 0.1$ ($\rm{r}\simeq400$ Mpc).  The magnitude and direction of the flow are claimed to be consistent with the peculiar velocity of the Local Group with respect to the CMB frame as inferred from the CMB dipole \citep{kogut93}.  Velocity coherence over such large scales is not predicted by the standard $\Lambda\rm{CDM}$ cosmology and would, if confirmed, constitute a major observational result.

In this paper we revisit the analysis presented in \cite{kashlinsky09}, hereafter referred to as K09.  The K09 analysis seeks to measure the kSZ signal of a sample of $\sim$700 X-ray-selected galaxy clusters.  The 3-year WMAP temperature maps for 8 differencing assemblies \citep{hinshaw07} are high-pass filtered in an attempt to remove the primary CMB anisotropy.  The temperatures of the filtered maps at the galaxy cluster locations are fit to a dipole, which is interpreted as the kSZ signal induced by a bulk flow of the galaxy clusters.

We will argue that the uncertainty of this measurement is dominated by primary CMB anisotropy, not detector noise.  As the CMB is observed by all 8 WMAP channels, the errors are highly correlated between these channels, and the inferred detection significance is greatly reduced.

\section{Cluster Sample}
\label{sec:clusters}
We construct a cluster sample as similar to that used in K09 as possible.  For all clusters we require that $z\le0.3$ and a corrected X-ray flux in the 0.1-2.4 keV band $> 3*10^{-12} \rm{erg} \rm{s}^{-1} \rm{cm}^{-2}$.  We use the REFLEX catalog \citep{boehringer04} and require that $\delta<2.5^{\circ}$, leaving 415 clusters.  We use the eBCS catalog \citep{ebeling98, ebeling00} and require that $\delta>0^{\circ}$ and $|b|>20^{\circ}$, leaving 279 clusters.  We use the CIZA catalog \citep{ebeling02,kocevski07}, which, after our cuts, contains 122 clusters.  Note that this version of the CIZA catalog contains 130 clusters total (73 from \cite{ebeling02} and 57 from \cite{kocevski07}), whereas K09 used an extended version containing 165 clusters total which is not publicly available at the time of this writing.  We note that while K09 removed all clusters whose X-ray emission appeared to be dominated by a point source, we do not.  This does not have a strong effect on our best-fit dipole, which, as discussed in Section \ref{sec:results}, is very close to the value presented in K09.  We use 816 clusters total, compared to 782 in K09.  720 of our clusters survive the 3-year WMAP KP0 galactic mask used in this analysis.

\section{CMB Maps}
\label{sec:cmb}
Our preparation of the filtered CMB maps duplicates the method used in K09.  We use the ``foreground reduced" temperature maps from the 3-year WMAP data release\footnote{http://lambda.gsfc.nasa.gov} \citep{hinshaw07}.  We use one map each from the 8 differencing assembly channels: Q1, Q2, V1, V2, W1, W2, W3, and W4.  For each channel we construct the filter $F_{\ell}$ described in K09.  The filter is essentially a high-pass filter with a transition multipole of $\ell\sim300$.  We explicitly remove the $\ell=0,1,2,3$ components from each map, as described in \cite{kashlinsky08}, by setting $F(\ell=0,1,2,3)=0$.  We use the 3-year WMAP KP0 galactic mask for all steps of the analysis which involve spherical harmonic transforms.  We use the HEALPix software package \citep{gorski05}.

The next step is to construct a mask which isolates the temperature fluctuations at the cluster locations.  This mask has all pixels outside of the cluster areas and outside of the KP0 galactic mask set to zero.  The ``cluster areas" are defined in K09 to be disc-shaped regions surrounding each cluster with $r_{\rm{disk}}=$min[6$\theta_{\rm{X-ray}},30^\prime$], where $\theta_{\rm{X-ray}}$ is related to each cluster's X-ray emission.  Because we lack access to the list of $\theta_{\rm{X-ray}}$ used in K09, we use $r_{\rm{disk}}=30^\prime$ for all clusters.  As pointed out in K09, because the majority of clusters have $\theta_{\rm{X-ray}}\ge5^\prime$, the K09 analysis uses $r_{\rm{disk}}=30^\prime$ for most clusters.  Specifically, the average $r_{\rm{disk}}$ used in K09 is $28^\prime.4$, the standard deviation is $3^\prime.2$, and only 16 clusters have $r_{\rm{disk}} < 20^\prime$.  As such, our method is very similar to the method used in K09.  As discussed in Section \ref{sec:results}, any difference in methodology has little impact on our best-fit dipole, which is very similar to that presented in K09.

The final step is to calculate the dipole component of the temperature fluctuations at the cluster locations (which have been isolated using the pixel mask described in the previous paragraph).  This is accomplished with the HEALPix function \textit{remove\_dipole}, which returns a 3-component dipole vector [$a_x,a_y,a_z$].  The dipole is calculated for each of the 8 WMAP channels individually and results from all channels are combined using inverse-variance weighting.  For example, $\hat{a}_x=(\sum_{i}w_{x,i}\hat{a}_{x, i})/ (\sum_{i}w_{x,i})$, where $w_{x,i}=\sigma_{x,i}^{-2}$ and $i$ is the index for the 8 WMAP channels.  The variances are calculated from simulations which are described in Section \ref{sec:errors}.

\section{Results}
\label{sec:results}

Our best-fit dipole is [$a_x,a_y,a_z$]=[1.2, -2.4, 0.2] $\mu$K.  This may be compared to [$a_x,a_y,a_z$]=[0.6, -2.7, 0.6] $\mu$K, the corresponding all-$z$ result from K09.  The magnitude (direction) of our best-fit dipole is within 6\% ($17^{\circ}$) of the best-fit dipole measured in K09.  We conclude that our best-fit dipole agrees well with that measured in K09, suggesting that the slight differences in methodology (the cluster catalogs and the choice of $r_{\rm{disk}}$) are unimportant.  Furthermore, we have repeated this analysis with $r_{\rm{disk}}$ increased and decreased by 20\%, and the results do not change significantly.  The magnitude (direction) of the best-fit dipole changes by less than 4\% ($11^\circ$) compared to the $r_{\rm{disk}}=30^\prime$ case.

Our best-fit dipole may be described as a vector with $2.7 \mu$K magnitude and with a higher temperature in the direction of the galactic coordinates $(\ell, b)=(298,4)$.  A naive interpretation of this dipole suggests a bulk flow moving away from $(\ell, b)=(298,4)$, which is the opposite sign of the velocity presented in K09.  However, the interpretation of the sign and magnitude of the velocity is complicated by the convolution of the kernel of $F_\ell$ with the gas profiles assumed for the clusters.  This issue was not discussed in K09.  For these reasons we do not attempt to quantify the sign or magnitude of the inferred velocity.  We feel that this is justified by the low detection significance of the dipole, as discussed in Section \ref{sec:detection}.

\section{Error Estimation}
\label{sec:errors}

We estimate the error of the dipole measurement as follows.  The basic strategy is to repeat the analysis described in Section \ref{sec:cmb}, but with simulated WMAP data replacing actual WMAP data.  We generate 1000 realizations of the CMB using the best-fit 3-year WMAP $\Lambda$CDM $C_{\ell}^{TT}$ power spectrum.  We use 3-year results, as opposed to 5-year, to remain consistent with the analysis of K09.  We convolve each CMB realization with the beams of the 8 WMAP channels in order to simulate noise-free observations.  For each CMB realization we also generate white noise maps for each WMAP channel.  The noise maps are generated using the prescription outlined in the WMAP Three-Year Explanatory Supplement \citep{limon07}, with the noise variance in a given pixel inversely proportional to the number of times that pixel was observed.  Our simulated noise is uncorrelated between WMAP channels and between map pixels.  While the latter assumption is not strictly true for WMAP data, the effects are negligible for any temperature analysis \citep{limon07}.  The white noise maps are added to the simulated noise-free CMB maps to produce maps which should have the same statistical properties as the actual ``foreground reduced" temperature maps described in Section \ref{sec:cmb}.  We have confirmed that the (KP0-masked) power spectra of these simulated maps match those of the true maps.

These simulated maps are then analyzed using the methods described in Section \ref{sec:cmb}.  To summarize, the maps are filtered by $F_{\ell}$ and the dipole component of the filtered maps at the locations of the galaxy clusters is calculated.  Although the WMAP data is simulated, we use the actual cluster positions and the actual KP0 galactic mask.  We note that because the definition of $F_{\ell}$ given in K09 depends on the measured spectrum $C_{\ell}^{sky}$, we calculate $F_{\ell}$ for each simulated map.

\begin{figure*}
\begin{center}
\includegraphics[width=1.0\textwidth]{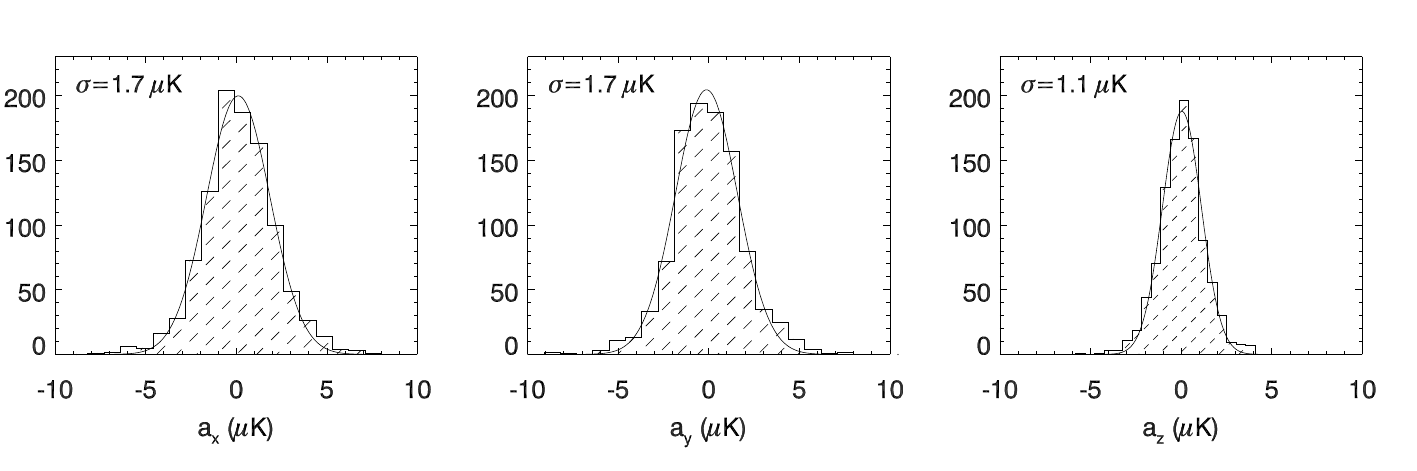}
\caption{1000 simulated estimates for the 3 dipole components.  These simulations take into account the CMB correlations between the different WMAP channels.  The uncertainty is highest on the dipole components that lie in the galactic plane ($a_x$ and $a_y$) because of the geometry of the galactic mask.}
\label{fig:err}
\end{center}
\end{figure*}

This suite of 1000 simulated WMAP ``experiments'' allows us to measure the uncertainty with which each WMAP channel measures each dipole component.  As there is only ``noise'' (CMB and detector noise) in these simulations, the distribution of dipole measurements provides a measure of the uncertainty of a single measurement of the dipole.  These distributions are well described as Gaussian with zero mean.  Inverse-variance weighting is used to combine the different channels' estimates of each dipole component.  Finally we are left with 1000 measurements of the 3 dipole components.  These distributions are also well described as Gaussian with zero mean and are shown in Figure \ref{fig:err}.  The $1\sigma$ measurement error for each dipole component is defined to be the best-fit Gaussian width $\sigma$ of each distribution.  The final uncertainties on the dipole components are [$\sigma_{a_x}, \sigma_{a_y}, \sigma_{a_z}$]=[1.7, 1.7, 1.1] $\mu$K.  These uncertainties should be accurate to within a few percent.  The uncertainty is highest on the dipole components that lie in the galactic plane ($a_x$ and $a_y$) because of the geometry of the galactic mask.

\section{CMB Correlations}
\label{sec:correlations}

\begin{figure*}
\begin{center}
\includegraphics[width=0.7\textwidth]{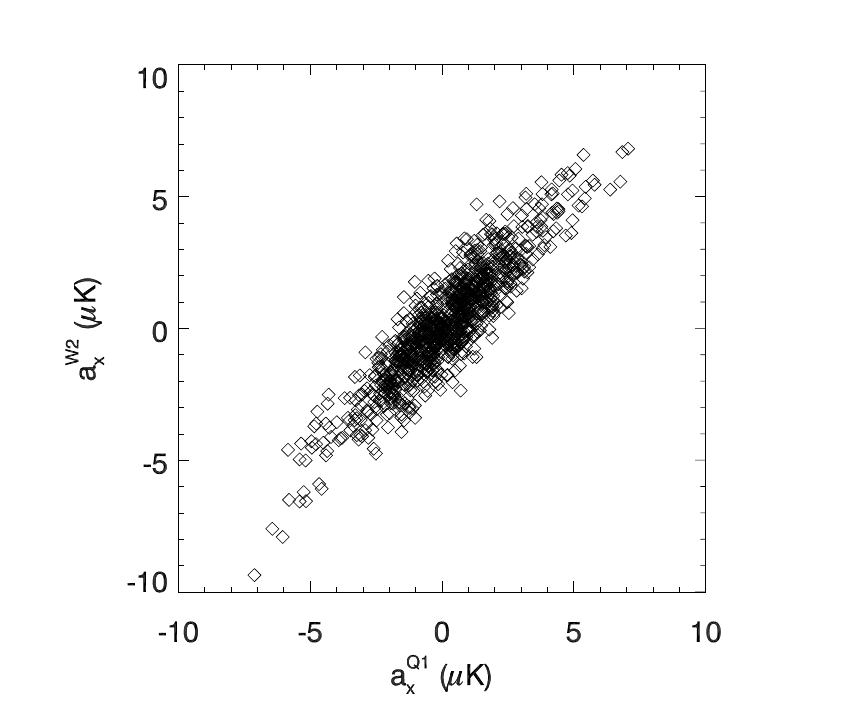}
\caption{1000 simulated estimates for the $a_x$ dipole component from two randomly chosen WMAP channels, Q1 and W2.  The estimates are highly correlated ($\rho=0.9$).  This high level of correlation is common to all pairs of channels and is caused by primary CMB fluctuations.}
\label{fig:corr}
\end{center}
\end{figure*}

These simulations bring to light an important fact that was not discussed in K09: the dipole estimates are highly correlated between the 8 different WMAP channels.  The correlation coefficient $\rho$ between two different channels' estimates of a given dipole component is approximately $0.9$.  For example, the correlation coefficient $\rho$ between the estimates of $a_x$ provided by the Q1 and W2 channels is 0.90, as shown in Figure \ref{fig:corr}.  We attribute these correlations to primary CMB fluctuations, as the detector noise is uncorrelated between channels.

To test the hypothesis that these correlations are caused by CMB fluctuations, we have separately filtered the noise-free CMB maps and the detector-noise-only maps in these 1000 experiments.  The filter $F_\ell$ is still constructed using the full maps which include the CMB and detector noise.  The resulting uncertainties on the dipole components in the CMB-only maps are [$\sigma_{a_x}, \sigma_{a_y}, \sigma_{a_z}$]=[1.7, 1.7, 1.1] $\mu$K.  These uncertainties are consistent with those obtained using the full simulations which contain both CMB and detector noise.  The uncertainties on the dipole components in the noise-only maps are much smaller: [$\sigma_{a_x}, \sigma_{a_y}, \sigma_{a_z}$]=[0.2, 0.2, 0.1] $\mu$K.  We conclude that the uncertainty of the dipole measurement is dominated by CMB fluctuations, not detector noise.  Although the filter $F_\ell$ is constructed with the intent of filtering out primary CMB anisotropy, residual CMB power is still present in the filtered maps and dominates the error on the dipole measurement.

Furthermore, if CMB fluctuations dominate the uncertainty of the dipole measurement, then the 5-year WMAP maps should produce a best-fit dipole that is very similar to that obtained using the 3-year WMAP maps.  We have repeated our analysis on 5-year WMAP maps\footnotemark[1] \citep{hinshaw09} to test this hypothesis.  We use the 3-year KP0 mask and 3-year $F_{\ell}$ filters on the 5-year maps in order to make the comparison as direct as possible.  The best-fit dipole is [$a_x,a_y,a_z$]=[1.3, -2.3, 0.1] $\mu$K, which is very close to the best-fit dipole from the 3-year WMAP data.  The magnitude (direction) of this dipole is within 2\% ($4^\circ$) of that obtained using the 3-year WMAP maps.  This provides further support to the claim that the uncertainty of the dipole measurement is dominated by CMB fluctuations, not detector noise.

If we simulate 1000 WMAP ``experiments'' and enforce the unphysical condition that each WMAP channel observes a different realization of the CMB, then the dipole estimates are uncorrelated, as expected.  In this unrealistic scenario the uncertainty on the dipole measurement is much smaller: [$\sigma_{a_x}, \sigma_{a_y}, \sigma_{a_z}$]=[0.7, 0.7, 0.4] $\mu$K.  These errors are closer to those presented in K09: [$\sigma_{a_x}, \sigma_{a_y}, \sigma_{a_z}$]=[0.5, 0.4, 0.4] $\mu$K.  These errors are smaller than the errors described above (which account for CMB correlations) by a factor of $\sim\sqrt{8}$, as is expected when combining 8 uncorrelated estimates as opposed to combining 8 highly correlated estimates.

\section{Detection Significance}
\label{sec:detection}

Our best-fit dipole is [$a_x,a_y,a_z]=[1.2 \pm 1.7, -2.4 \pm 1.7, 0.2 \pm 1.1]$ $\mu$K.  The $\chi^2$/d.o.f. is 2.52/3.  The probability to exceed this $\chi^2$ is 0.47, corresponding to a Gaussian detection significance of 0.7$\sigma$.  If we use the best-fit dipole presented in K09, [$a_x,a_y,a_z$]=[0.6, -2.7, 0.6] $\mu$K, which is quite close to ours, the detection significance is 0.8$\sigma$.  The errors used in these significance calculations come from the simulations presented in Section \ref{sec:errors}, which take into account correlations between the WMAP channels.

A slightly different statistic may be considered if one is specifically interested in constraining bulk flows: the component of the best-fit dipole projected along the direction of the peculiar velocity of the Local Group with respect to the CMB frame.  Measurements of the CMB dipole  \citep{kogut93} suggest that this velocity is towards the galactic coordinates $(\ell, b)=(276,30)$.  This statistic has 1 degree of freedom and we have calculated the uncertainty on its measurement using the methods described in Section \ref{sec:errors}.  The best-fit projected dipole is 2.2 $\pm$1.6 $\mu$K, corresponding to a detection significance of 1.4$\sigma$.  The sign is such that the temperature is higher at $(\ell, b)=(276,30)$.

We conclude that there is not a significant detection of a bulk flow.  The significance of the best-fit bulk flow is $0.7\sigma$ and the significance of the component projected along the Local Group's peculiar velocity is $1.4\sigma$.

\section{Conclusion}
\label{sec:conclusion}

We have revisited the analysis presented in \cite{kashlinsky08, kashlinsky09} which reports a significant detection of a bulk flow of $\sim$700 galaxy clusters out to $z\simeq0.1$ by means of the kSZ effect.  We have demonstrated that the estimates for the kSZ signal are highly correlated between the different WMAP channels used in this analysis and that this correlation is caused by primary CMB anisotropy.  We have simulated the errors on the kSZ measurement while taking into account these CMB correlations and find that there is not a significant detection of a kSZ signal or bulk flow.


\acknowledgments
The author would like to thank Bradford Benson, John Carlstrom, Tom Crawford, William Holzapfel, Wayne Hu, Stephan Meyer, and Christian Reichardt for useful discussions.

This work was supported by the NSF Physics Frontier Center award PHY-0551142 and the NSF OPP award ANT-0638937.

\end{document}